\documentclass[epj]{svjour}
\RequirePackage{float}
\usepackage{epstopdf}
\usepackage{hyperref}
\usepackage{amssymb}
\usepackage{mathtools}
\usepackage{capt-of}
\usepackage{graphicx}  
\usepackage{dcolumn}   
\usepackage{bm}%

\begin{document}
\title{Big Bang Nucleosynthesis and Entropy Evolution in $f(R,T)$ Gravitation}
\author{Snehasish Bhattacharjee\inst{1} \and P.K. Sahoo\inst{2}
}                     
%
%
\institute{Department of Astronomy, Osmania University, Hyderabad-500007,
India,  Email: snehasish.bhattacharjee.666@gmail.com \and Department of Mathematics, Birla Institute of
Technology and Science-Pilani, Hyderabad Campus, Hyderabad-500078,
India,  Email:  pksahoo@hyderabad.bits-pilani.ac.in}
\date{Received: 13th Jan 2020 / 24th March 2020 }
%
\abstract{
The present article is devoted to constrain the model parameter $\chi$ for the $f(R,T)= R + \chi T$ gravity model by employing the constraints coming from big bang nucleosynthesis. We solve the field equations and constrain $\chi$ in the range $-0.14 \kappa^{2} \leq \chi \leq 0.84 \kappa^{2}$ (where $\kappa^{2} = \frac{8 \pi G}{c^{4}}$) from the primordial abundances of light elements such as helium-4, deuterium and lithium-7. We found the abundances of helium-4 and deuterium agrees with theoretical predictions, however the lithium problem persists for the $f(R,T)$ gravity model. We also investigate the evolution of entropy for the constrained parameter space of $\chi$ for the radiation and dust universe. We report that entropy is constant when $\chi = 0$ for the radiation dominated universe, whereas for the dust universe, entropy increases with time. We finally use the constraints to show that $\chi$ has negligible influence on the cold dark matter annihilation cross section.
} 
\titlerunning{Big Bang Nucleosynthesis and Entropy Evolution in $f(R,T)$ Gravitation} \authorrunning{S. Bhattacharjee, P.K. Sahoo}
\maketitle
\section{Introduction}
The current accelerated expansion of the universe favors big bang cosmology. The model predicts the abundances of several light elements of the primordial universe with great precision. The elements were produced as a result of nuclear fusion started seconds after the big bang and lasted for some minutes. Additionally, the model predicts inflation which is a super exponential increase of the volume of the universe for a very short time ($10^{-43}$ sec). Inflation have been successful in solving the flatness, horizon and homogeneity problems of the universe \cite{1}. \\
However, many cosmological puzzles exist which hitherto cannot be explained by the standard big bang cosmology such as origin of dark matter and dark energy, cosmological constant problem, cosmic coincidence problem and the exact form of the inflation potential etc. \cite{theory,2,3}. To answer these problems, modifying GR have become a promising alternative giving rise to a plethora of modified gravity theories. \\
$f(R,T)$ gravity is a widely studied modified gravity theory introduced in the literature in \cite{harko} and is a generalization of $f(R)$ gravity (see \cite{extended} for a review on modified gravity theories). In this theory, the Ricci scalar $R$ in the action is replaced by a combined function of $R$ and $T$ where $T$ is the trace of the energy-momentum tensor. $f(R,T)$ gravity have been widely employed in various cosmological scenarios and have yielded interesting results in areas such as dark matter \cite{in22} dark energy \cite{in21}, super-Chandrasekhar white dwarfs \cite{in25}, massive pulsars \cite{in23}, wormholes \cite{in26}, gravitational waves \cite{in36}, baryogenesis \cite{baryo}, bouncing cosmology \cite{bounce} and in varying speed of light scenarios \cite{physical}.\\
In this article we are interested in constraining the model parameter of $f(R,T)$ gravity theory for the ansatz $f(R,T) = R + \chi T$, where $\chi$ is the model parameter. Constraining $\chi$ can help us to better understand the impact of $\chi$ in cosmological models and also in the above mentioned astrophysical areas.\\
Big Bang nucleosynthesis can be an excellent way to constrain the model parameters of any modified gravity theory as the abundances of primordial light elements such as deuterium ($^{2}H$), helium ($^{4}He$) and lithium ($^{7}Li$) have been observationally constrained to great accuracy. These abundances are directly related to the Hubble parameter $H$, which ultimately involve the model parameters of any chosen modified gravity theory. This method have been successfully employed to constrain the model parameters of $f(R)$ gravity \cite{cross,avelino}, $f(\mathcal{T})$ gravity \cite{torsion}, scalar-tensor gravity models \cite{coc} and to test the viabilities of Brans Dicke cosmology with varying $\Lambda$ \cite{brans}, Higher Dimensional Dilaton Gravity Theory of Steady-State Cosmological (HDGS) model in the context of string theory \cite{theory} and massive gravity theory \cite{massive}. The discrepancy between predicted and observed abundance of lithium ('lithium problem') is investigated in \cite{lithium} (and in references therein).\\
The paper is organized as follows: In Section \ref{sec2} we provide an overview of $f(R,T)$ gravity. In Section \ref{sec3} we summarize big bang nucleosynthesis and present a through analysis to constrain $\chi$. In Section \ref{sec5}, we investigate the evolution of entropy for the radiation and matter filled universes for the constrained range of $\chi$. In Section \ref{sec7}, we investigate whether $\chi$ influences the cold dark matter annihilation cross section and Section \ref{sec6} is devoted to discussions and conclusions.  

\section{Overview of $f(R,T)$ Gravity}\label{sec2}

The action in $f(R,T)$ gravity is given by 
\begin{equation}\label{1}
\mathcal{S}= \int \sqrt{-g}\left[\frac{1}{2 \kappa^{2}} f(R,T) + \mathcal{L}_{m}\right] d^{4}x
\end{equation} 
where $\mathcal{L}_{m}$ represent matter Lagrangian and $\kappa^{2} = \frac{8 \pi G}{c^{4}}$.\\
Stress-energy-momentum tensor for the matter fields is given as 
\begin{equation}\label{2}
T_{\mu \nu} = \frac{-2}{\sqrt{-g}}\frac{\delta (\sqrt{-g} \mathcal{L}_{m} )}{\delta g^{\mu \nu}}=g_{\mu \nu}\mathcal{L}_{m}-2\frac{\delta \mathcal{L}_{m} }{\delta g^{\mu \nu}}
\end{equation}
varying the action (\ref{1}) with respect to the metric yields 
\begin{equation}\label{3}
f^{1}_{,R}(R,T)R_{\mu \nu}+\Pi_{\mu \nu}f^{1}_{,R}(R,T) -\frac{1}{2}g_{\mu\nu}f(R,T) =\kappa^{2}T_{\mu \nu}-(T_{\mu \nu}+\Theta_{\mu \nu})f^{1}_{,T}(R,T)  
\end{equation}
where 
\begin{equation}\label{4}
 - \nabla_{\mu}  \nabla_{\nu}+g_{\mu \nu} \square=\Pi_{\mu \nu}
\end{equation}
\begin{equation}\label{5}
g^{\alpha \beta}\frac{\delta T_{\alpha \beta}}{\delta g^{\mu \nu}}\equiv \Theta_{\mu \nu}
\end{equation}
and $f^{i}_{,X}\equiv \frac{d^{i}f}{d X^{i}}$. 
Upon contraction (\ref{3}) with $g^{\mu \nu}$, the trace of the field equations is obtained as 
\begin{equation}\label{6}
f^{1}_{,R}(R,T)R -2f(R,T)+3\square f^{1}_{,R}(R,T)= -  (\Theta+T)f^{1}_{,T}(R,T)+\kappa^{2}T 
\end{equation}
We now consider a flat FLRW background metric as \begin{equation}\label{7}
ds^{2}=dt^{2}-a(t)^{2}[dx^{2}+dy^{2}+dz^{2}]
\end{equation}
where $a(t)$ denote the scale factor. For a universe dominated by a perfect fluid the matter Lagrangian density is given as $\mathcal{L}_{m}=-p$. Upon employing this to (\ref{3}) and (\ref{6}) yields \begin{equation}\label{8}
\frac{1}{f^{1}_{,R}(R,T)}\left[ -3 \dot{R} H  f^{2}_{,R}(R,T)+pf^{1}_{,T}(R,T)-Rf^{1}_{,R}(R,T) + \frac{1}{2}\left( f(R,T)\right)  \right]+\frac{f^{1}_{,T}(R,T)+\kappa^{2}}{f^{1}_{,R}(R,T)} \rho=3H^{2}
\end{equation}
\begin{multline}\label{9} 
 \frac{1}{f^{1}_{,R}(R,T)} \left[  -\frac{1}{2}\left( f(R,T)+\dot{R}^{2} f^{3}_{,R}(R,T)+\ddot{R} f^{2}_{,R}(R,T) -R f^{1}_{,R}(R,T)\right)  -pf^{1}_{,T}(R,T) + 2H \dot{R}f^{1}_{,R}(R,T) \right]\\ + \frac{f^{1}_{,T}(R,T)+\kappa^{2}}{f^{1}_{,R}(R,T)}p =-3H^{2} -2\dot{H} 
\end{multline}
where $H$ denote the Hubble parameter, overhead dots denote derivative with respect to time,  $p$ represents pressure and $\rho$ represents density with $T=\rho - 3p$.\\
setting $f(R,T)$ functional form to be 
\begin{equation}\label{10}
f(R,T) = R + \chi T.
\end{equation}
Substituting \eqref{10} in \eqref{8} and solving for Hubble parameter ($H_{f(R,T)}$), we obtain  
\begin{equation}\label{11}
H_{f(R,T)}=\frac{\epsilon}{t},
\end{equation}
where

\begin{equation}\label{12}
\epsilon = -\frac{1}{3}\left[ \frac{ - 2\kappa^{2} +\left( \omega - 3  \right)\chi  }{\left( \omega+1\right)\left( \chi+\kappa^{2}  \right)   }\right] .
\end{equation}
where $\omega = p/\rho$ denote the EoS parameter. 
The scale factor $a(t)$ takes the form 
\begin{equation}
a \sim t^{\epsilon}
\end{equation}
The expression of density $\rho$ reads 
\begin{equation}
\rho = \frac{\left( 2 \kappa^{2} - \chi (\omega - 3)\right)\left(\chi \left(3 + 8 \chi - \omega \right) + 2 \kappa^{2} \left(1 + \chi \left(3 + \omega \right)  \right)   \right)  }{3 t^{2}\left(\kappa^{2} + \chi \right)^{2}\left( 1 + 6 \chi + 8 \chi ^{2}\right)\left(1 + \omega \right) ^{2}  }
\end{equation}
For a radiation dominated universe ($\omega = 1/3$), the expression of Hubble parameter reads
\begin{equation}\label{14}
H_{f(R,T)} = \left[ \frac{8 \chi/3 + 2 \kappa^{2}}{4 (\kappa^{2} + \chi)}\right]/t 
\end{equation}
In Einstein's GR, the expression of Hubble parameter in radiation dominated universe reads
\begin{equation}\label{13}
H = \frac{1}{2 t}
\end{equation}

\section{Nucleosynthesis in $f(R,T)$ gravity}\label{sec3}

In this method we are interested in finding a suitable value or range of $\chi$ which can suffice the primordial abundances of light elements. Specifically we will be studying the ratio of Hubble parameter in $f(R,T)$ gravity to the Hubble parameter of standard big bang cosmology for the radiation dominated universe. The ratio is represented as
\begin{equation}\label{15}
Z=\frac{H_{f(R,T)}}{H_{SBBN}}
\end{equation}
where $H_{f(R,T)}$ is given by \eqref{14} and $H_{SBBN}$ is given by \eqref{13} and SBBN stands for Standard Big Bang nucleosynthesis. The primordial abundances of the light elements ($^{2}D$, $^{4}He$, $^{7}Li$) depend on the expansion rate of the universe and on the baryon density \cite{epjp52,epjp53}. The baryon density parameter reads \begin{equation}
\eta_{10} \equiv 10^{10}\eta_{B}\equiv 10^{10} \frac{\eta_{B}}{\eta_{\gamma}}
\end{equation}
Where $\eta_{10} \simeq 6 $ and $\eta_{B}$ represents the baryon to photon ratio \cite{epjp56}.\\
$Z\neq 1$ correspond to non-standard expansion factor. This can arise due to GR modification or due to the presence of additional light particles such as neutrinos which would make the ratio to be, $Z = \left( 1 + \frac{7}{43} (N_{\nu} - 3)\right)^{1/2} $ \cite{theory}. However, we are interested for the case where the value of $(Z-1)$ comes from GR modification and hence we shall assume $N_{\nu} = 3$.

\subsection{$^{4}He$ abundance in $f(R,T)$ Gravity}

The first step in producing helium ($^{4}He$) starts with producing $^{2}H$ from a neutron ($n$) and a proton ($p$). After that, Deuterium is converted into $^{3}He$ and Tritium ($T$).
\begin{equation}
n + p \rightarrow ^{2}H + \gamma; \hspace{0.25in} ^{2}H+^{2}H \rightarrow ^{3}He + n; \hspace{0.25in} ^{2}H + ^{2}H \rightarrow ^{3}H + p
\end{equation}
$^{4}He$ is finally produced from the combination of $^{3}H$ with $^{2}H$ and $^{3}He$;
\begin{equation}
^{2}H + ^{3}H \rightarrow ^{4}He + n; \hspace{0.4in} ^{2}H + ^{3}He\rightarrow ^{4}He + p
\end{equation}
The simplest way to ascertain the $^{4}He$ abundance is from the numerical best fit given in \cite{epjp57,epjp58}
\begin{equation}
Y_{p} = 0.2485 \pm 0.0006 + 0.0016 \left[\left( \eta_{ 10} - 6\right) +100\left( Z-1\right)  \right] 
\end{equation}
For $Z=1$, we recover the SBBN $^{4}He$ fraction, which reads $(Y_{p})|_{SBBN} = 0.2485 \pm 0.0006$.\\
Observations reveal the $^{4}He$ abundance to be $0.2449 \pm 0.0040$ \cite{jcap2020}. Thus we obtain 
\begin{equation}
0.2449 \pm 0.0040 = 0.2485 \pm 0.0006 +0.0016 \left[100(Z-1) \right] 
\end{equation}
where we have set $\eta_{ 10} = 6$. This constrains $Z$ in the range  $1.0475 \pm 0.105$. 

\subsection{$^{2}H$ abundance in $f(R,T)$ Gravity}

Deuterium $^{2}H$ is produced from the reaction $n+p \rightarrow ^{2}H +\gamma$. Deuterium abundance can be ascertained from the numerical best fit given in \cite{epjp52} 
\begin{equation}\label{16}
y_{D p} = 2.6(1 \pm 0.06) \left( \frac{6}{\eta_{ 10} - 6 (Z-1)}\right)^{1.6} 
\end{equation}
For $Z=1 \& \eta_{ 10} = 6$, $y_{D p} |_{SBBN} = 2.6 \pm 0.16$. Observational constraint on deuterium abundance is $y_{D p} = 2.55 \pm 0.03$ \cite{jcap2020}. Thus equating this to \ref{16}, we obtain 
\begin{equation}
2.55 \pm 0.03 = 2.6(1 \pm 0.06) \left( \frac{6}{\eta_{ 10} - 6 (Z-1)}\right)^{1.6} 
\end{equation}
This constraints $Z$ in the range $Z=1.062 \pm 0.444$. The constraint on $Z$ for the deuterium abundance partially overlaps with that of the helium abundance. Thus $\chi$ can be fine tuned to fit the abundances for both $^{2}H$ and $^{4}He$. 

\subsection{$^{7}Li$ abundance in $f(R,T)$ Gravity }  
 
The lithium abundance is puzzling in the sense that the $\eta_{ 10}$ parameter which precisely fits the abundances of other elements successfully does not fit the observations of $^{7}Li$ and the ratio of the expected SBBN value of $^{7}Li$ abundance to the observed one is between $2.4-4.3$ \cite{theory,theory42}. Thus neither SBBN nor any modified gravity theory can suffice the low abundance of $^{7}Li$. This is known as the Lithium problem \cite{theory}.\\
The numerical best fit expression for $^{7}Li$ abundance reads \cite{epjp52}
\begin{equation}
y_{Li p} = 4.82 (1 \pm 0.1)\left[\frac{\eta_{ 10} - 3 (Z-1)}{6} \right] ^{2}
\end{equation}
Observational constraint on lithium abundance is $y_{Li p} = 1.6 \pm 0.3$ \cite{jcap2020}. The constraint on $Z$ to fit the $^{7}Li$ abundance is $Z = 1.960025 \pm 0.076675$ which clearly does not overlap with the deuterium-2 and helium-4 constraints. 

\subsection{Results}

\begin{figure}[H]
  \centering
  \includegraphics[width=8.5cm]{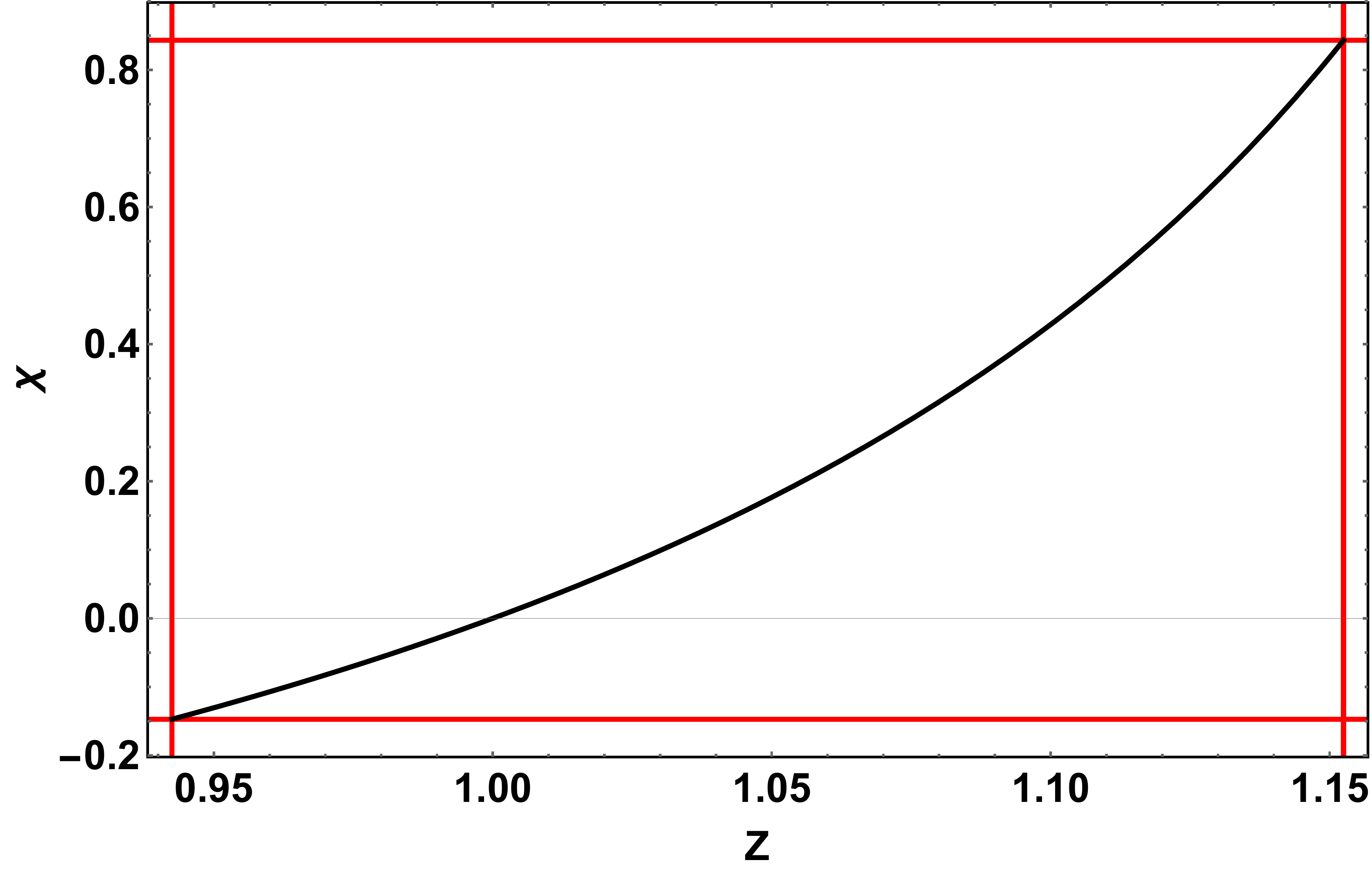}
  \caption{$\chi$ as a function of $Z$. Vertical lines are drawn at $Z=0.9425$ and $Z=1.1525$ while the horizontal lines are drawn at $\chi = -0.14$ and $ \chi = 0.84$.}
  \label{1st}
\end{figure}

From Table \ref{table} it is clear that $f(R,T)$ gravity yields excellent estimates for the abundances of helium and deuterium which match better to observations than the SBBN model. However, the abundance of lithium is still a problem for both the models (SBBN and $f(R,T)$ gravity). In Figure \ref{1st}, we show $\chi$ as a function of $Z$. For $Z$ in the range $0.9425\leq Z\leq 1.1525$, the theoretical predictions for the abundances of deuterium and helium agrees with observations. This constraints $\chi$ in the range $-0.14 \kappa^{2}\lesssim \chi \lesssim 0.84 \kappa^{2}$. 

\captionof{table}{The abundances He-4, Deuterium and Li-7 for different models
}\label{table}
\begingroup
\setlength{\tabcolsep}{10pt} 
\renewcommand{\arraystretch}{1.5} 
\begin{tabular}{ |p{5cm}||p{3.5cm}|p{3.5cm}|p{3.5cm}|  }
 \hline

 Models and data/Abundances    & $Y_{p}$ &$y_{Dp}$&$y_{Lip}$\\
 \hline
 SBBN model   & $0.2485 \pm 0.0006$    &$2.6 \pm 0.16$&   $4.82 \pm 0.48$\\
 $f(R,T)$ Gravity&   $0.2574 \pm 0.0006$  & $2.8485 \pm 0.1715$   &$5.29275 \pm 0.52925$\\
 Observational data &$0.2449 \pm 0.0040$ \cite{jcap2020} & $2.55 \pm 0.03$ \cite{jcap2020} & $1.6\pm 0.3$ \cite{jcap2020}\\

 \hline
 
\end{tabular}
\endgroup

\section{Entropy Evolution in $f(R,T)$ Gravity}\label{sec5}
\begin{figure}[H]
  \centering
  \includegraphics[width=10cm]{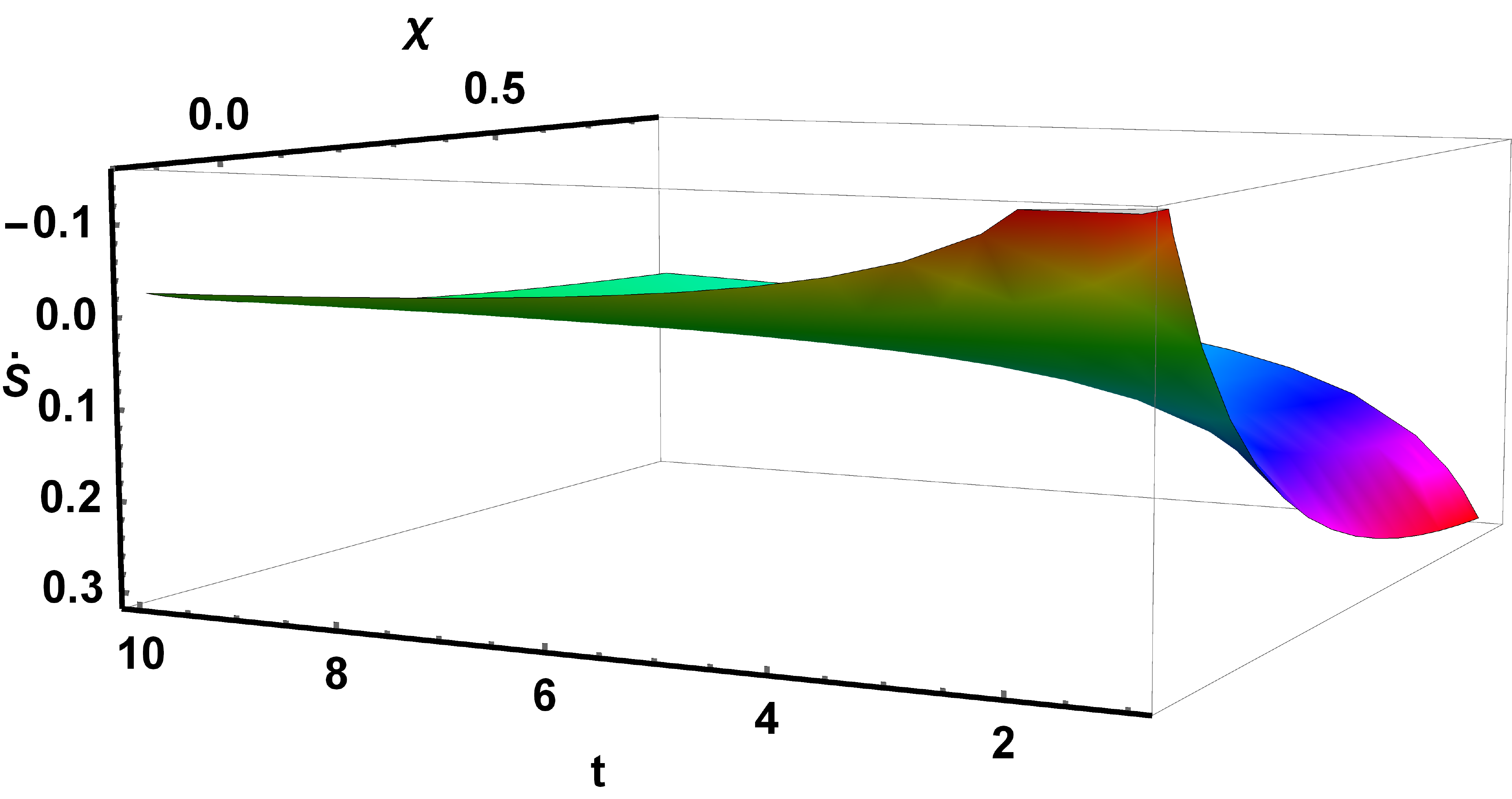}
  \caption{Time evolution of $\dot{S}$ in radiation universe for $-0.14 \kappa^{2}\lesssim \chi \lesssim 0.84 \kappa^{2}$.}
  \label{2nd}
\end{figure}

\begin{figure}[H]
  \centering
  \includegraphics[width=10cm]{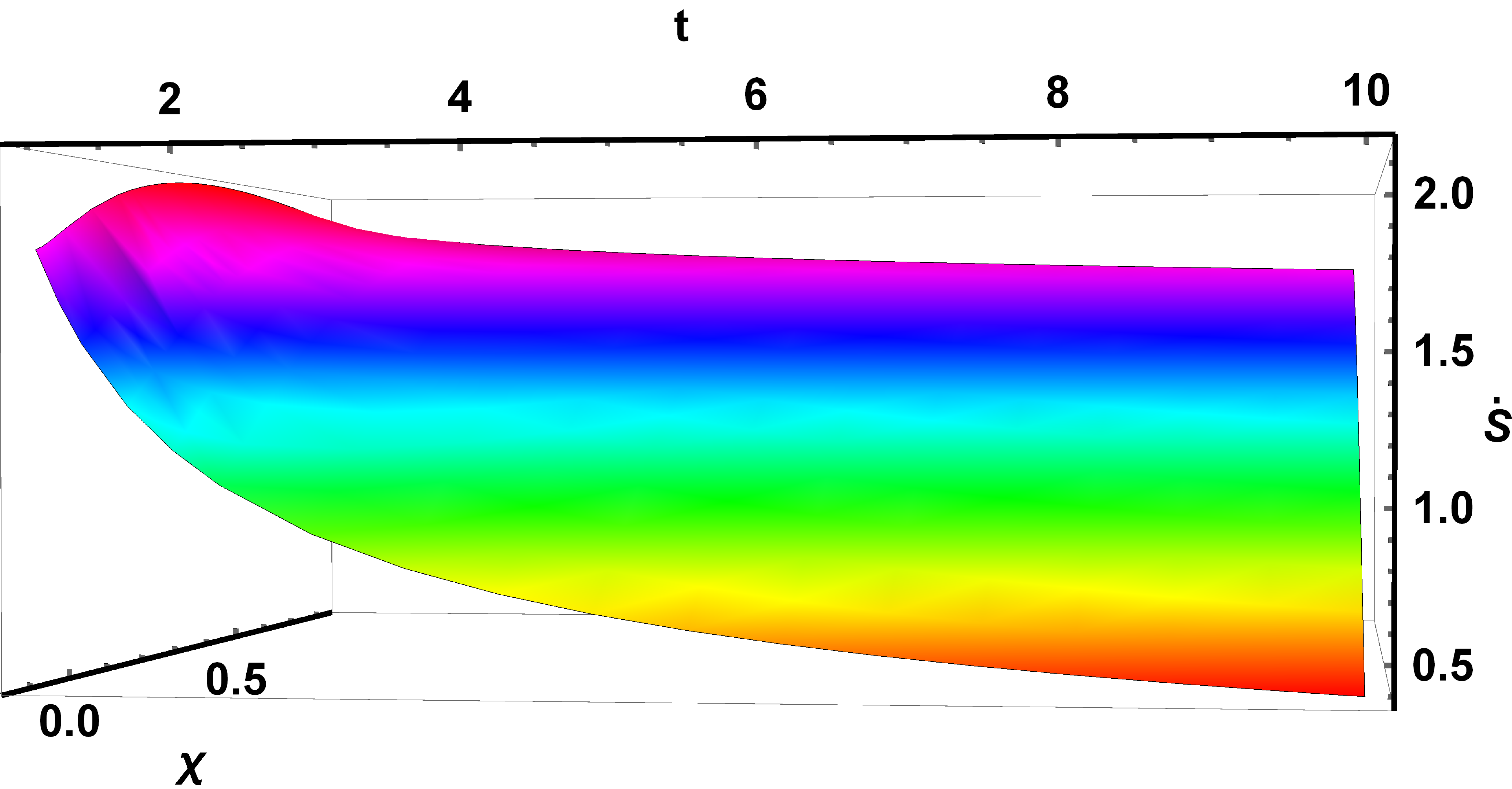}
  \caption{Time evolution of $\dot{S}$ in dust universe for $-0.14 \kappa^{2}\lesssim \chi \lesssim 0.84 \kappa^{2}$.}
  \label{3rd}
\end{figure}

Baryon to entropy ratio is a useful parameter characterizing the over abundance of matter over anti-matter in the universe. Since the law of conservation of energy momentum is not maintained in $f(R,T)$ gravity, we investigate how this affects adiabaticity \cite{bar17,bar}. In SBBN model, the entropy of the universe is a conserved quantity throughout its evolution and this is due to the fact that at low energies, baryon number is neither created nor destroyed since there are no decays and consequently the baryon to entropy ratio $\eta_{S}$ is a constant \cite{bar}. Equivalently, once the large scale annihilation processes have concluded, the baryon to photon ratio $\eta_{B}$ is also a constant, and both quantities can be connected easily \cite{bar}. \\
From the first law of thermodynamics, we obtain
\begin{equation}
dE + pdV = T dS
\end{equation}
where $S = s(a^{3})$ and $E=\rho (a^{3})$ are the entropy and internal energy of the universe respectively. This gives \cite{bar}
\begin{equation}
d (\rho a^{3}) + p d (a^{3}) = T d S \rightarrow \frac{T}{a^{3}} \dot{S} = \dot{\rho} + 4 H \rho
\end{equation} 
From statistical mechanics, density $\rho$ is related to temperature $T$ as \cite{turner}
\begin{equation}
\rho = \frac{\pi ^{2}}{30} g_{*s} T^{4}
\end{equation}
where $g_{*s} = 107$ is the effective number of relativistic degrees of freedom contributing to the entropy of the universe \cite{bar}.\\
Substituting all the values, we obtain 
\begin{equation}
\dot{S}= \frac{1.86121 t^{-\left(\frac{3 + 2 \chi}{2 + 2 \chi} \right) }\chi \left(0.15188 + 0.911281 \chi + 1.55255 \chi^{2} + 0.810028 \chi^{3} \right) }{(1+ \chi)^{3} \left(0.125 +0.75 \chi + \chi^{2} \right) \left[\frac{(3 + 4 \chi) (3 + 14 \chi + 12 \chi^{2})}{t^{2} (1 + \chi)^{2} ( 1 + 6 \chi + 8 \chi^{2})} \right]^{0.25} }
\end{equation} 
In Figure \ref{2nd}, we observe that $\dot{S}$ is positive for $\chi > 0$ and negative for $\chi < 0$ at early times but converges to zero at late times. From Table \ref{table2}, we further note that $\dot{S}=0$ for $\chi = 0$ (GR) for the radiation universe. However, Figure \ref{3rd} shows that $\dot{S}> 0$ for the dust universe and it is also evident that $\dot{S}$ increases as $\chi$ increases. It is also evident that $\dot{S}$ decreases slowly with time for the dust universe.

\caption{Rate of change of entropy ($\dot{S}$) for different models
}\label{table2}
\begingroup
\setlength{\tabcolsep}{10pt} 
\renewcommand{\arraystretch}{2.5} 
\begin{tabular}{ |p{3cm}||p{5cm}|p{5cm}|  }

 \hline
 Models & \multicolumn{2}{|c|}{Rate of change of entropy ($\dot{S}$)} \\
\cline{2-3}
 &Radiation universe ($\omega=1/3$) &Dust universe ($\omega=0$)\\
 \hline
GR   &   0 & $\left[ \frac{2.01018}{\frac{1}{t^{2}}^{0.25}t}\right] $\\
 \hline
 $f(R,T)$ Gravity&  $ \left[ \frac{-0.30863}{\frac{1}{t^{2}}^{0.25}t^{1.58625}}\right] \bigg |_{\chi =-0.14 }$  $\left[\frac{0.308305}{\frac{1}{t^{2}}^{0.25}t^{1.27125}}\right] \bigg |_{\chi = 0.84}$&$\left[ \frac{1.84325}{\frac{1}{t^{2}}^{0.25}t^{1.1725}}\right] \bigg |_{\chi =-0.14 }$  $\left[ \frac{2.05634}{\frac{1}{t^{2}}^{0.25}t^{0.5425}}\right] \bigg |_{\chi = 0.84}$\\
 \hline

\end{tabular}
\endgroup

\section{Dark Matter Annihilation Cross Section in $f(R,T)$ Gravity}\label{sec7}

Recent cosmological observations have constrained the normalized cold dark matter density in the range \cite{cross18}
\begin{equation}
0.075 \lesssim \Omega_{cdm}h^{2} \lesssim 0.126
\end{equation} 
In this section we shall assume dark matter to be composed of weakly-interacting massive particles (WIMPs). In \cite{cross} the authors derived an analytical expression where the WIMP cross section $\bar{\sigma}$ is written in terms of the relic density of dark matter, its mass $m$ and on the power $n$ for the power law $f(R)$ gravity model of the form $f(R) \sim R ^{n}$. We shall now investigate the role of $\chi$ in dark matter annihilation cross section for a given WIMP mass.\\
The expression relating the dark matter relic density, its mass, dark matter annihilation cross section and parameters of a modified gravity model reads \cite{cross}
\begin{equation}
\Omega_{cdm}h^{2} = 1.07 \times 10^{9} \frac{(\bar{m} + 1) x_{f}^{(\bar{m} + 1) GeV^{-1}}}{(h_{*}/g_{*s}^{1/2}) M_{p}\bar{\sigma}}
\end{equation}
where \begin{equation}
\bar{m} = m + \left(1 -  n  \right) 
\end{equation}
where $\bar{m} = m$  for GR and $m=0$ \& 1 correspond to s-wave and p-wave polarizations respectively and for $n = 1$, GR is recovered. \\
$x_{f}$ is the freeze-out temperature and given as \cite{turner,cross}
\begin{equation}
x_{f} = \ln [0.038 (\bar{m} + 1) (g / g_{*s}^{1/2}) M_{p} m \bar{\sigma}]  -(\bar{m} + 1) \ln[\ln [0.038 (\bar{m} + 1) (g / g_{*s}^{1/2}) M_{p} m \bar{\sigma}]]
\end{equation}
where $g=2$ is the spin polarizations of the dark matter particle \cite{cross} and $M_{p}$ is the Planck mass. \\
In \cite{cross}, the authors found substantial influence of $n$ in $\bar{\sigma}$ although $n$ had very little deviation from GR ($n- 1\lesssim 0.00016$). We now modify $\bar{m}$ for our $f(R,T)$  gravity model and check the influence of $\chi$ on $\bar{\sigma}$.\\
For $\bar{m}$ of the form 
\begin{equation}
\bar{m} = m + \chi
\end{equation}
we get $\bar{m}=m$ when $\chi = 0$. Now from BBN, $\chi$ is constrained in the range $-0.14 \kappa^{2}\lesssim \chi \lesssim 0.84 \kappa^{2}$ which is $\mathcal{O} 10^{-43}$. Hence our $f(R,T)$ gravity model produces $\bar{\sigma}$ very close to that predicted from GR. Nonetheless, it would be interesting to do the same analysis with $f(R,T)$ gravity models with a power-law dependence on $T$. 
  
\section{Discussions}\label{sec6}

Modified gravity theories are becoming popular owing to the failures of GR in explaining the current acceleration of the universe. In modified gravity theories, the model parameters are fine tuned to obtain the desired results which sometimes differ significantly from GR. In this work we investigate the viability of the most widely studied and simplest minimal matter-geometry coupled $f(R,T)$ gravity model of the form $f(R,T) = R + \chi T$ in cosmological models and in many astrophysical areas.\\ 
The present manuscript uses the constraints of abundances of light elements such as helium-4, deuterium and lithium-7 to constrain the model parameter $\chi$ to unprecedented accuracy. From the analysis, we report a tight constraint on $\chi$ in the range $-0.14 \kappa^{2}\lesssim \chi \lesssim 0.84 \kappa^{2}$.\\
We also study the evolution of entropy for the constrained parameter  space of $\chi$ for the radiation and dust universe. We report that entropy ($S$) is constant for $\chi = 0$ for the radiation dominated universe, whereas for the dust universe, $\dot{S} > 0$ for the allowed range of $\chi$.\\ 
We also found that $\chi$ has negligible influence on dark matter annihilation cross section ($\bar{\sigma}$) and produces $\bar{\sigma}$ very close to that predicted by GR.\\
The constraints on $\chi$ obtained from the present analysis makes it clear-cut that the parameter $\chi$ has negligible influence in cosmological models and in above mentioned astrophysical areas. It would certainly be interesting to apply the method to constrain the model parameters for other $f(R,T)$ gravity models and to check their viability in representing the current state of the universe.\\
As a final note we add that in \cite{pseudotensor}, the authors reported that the gravitational energy-momentum pseudotensor can also be an important tool in distinguishing and constraining different theories of gravity. Specifically, in \cite{52pseudo}, the authors reported that the gravitational pseudotensor is useful to identify the dissimilarities in quadrupolar gravitational radiation coming from Einstein's gravity and $f(R)$ gravity. This idea was further extended to teleparallel gravity in \cite{pseudotensor}. Since gravitational waves differ substantially from one theory of gravity to another \cite{53pseudo}, detection of the polarization modes of the gravitational radiation can be promising to constrain extended theories of gravity \cite{pseudotensor}.  

\section*{Acknowledgments}
SB thank Biswajit Pandey for constant support and motivation. SB also thank Suman Sarkar and Biswajit Das for helpful discussions. PKS acknowledges CSIR, New Delhi, India for financial support to carry out the Research project [No.03(1454)/19/EMR-II Dt.02/08/2019]. We are very much grateful to the honorable referee and the editor for the illuminating suggestions that have significantly improved our work in terms of research quality and presentation.

\end{document}